\documentstyle[12pt]{article}

\large
\newcommand{\be}{\begin{eqnarray}}
\newcommand{\ee}{\end{eqnarray}}
\newcommand\del{\partial}

\begin{document}
\setlength{\baselineskip}{21pt}
\pagestyle{empty}
\vfill
\eject
\begin{flushright}
SUNY-NTG-94/18 \\
TPI-MINN-94/10-T
\end{flushright}

\vskip 1.0cm
\begin{center}
{\bf Spectral sum rules and finite volume partition function in gauge
theories with real and pseudoreal fermions.}
\end{center}
\vskip 1.0 cm
\centerline{A. Smilga }
\vskip .2cm
\centerline{Institute for Theoretical Physics}
\centerline{University of Minnesota, Minneapolis MN 55455
\footnote{On leave of absence from ITEP, B. Cheremushkinskaya 25, Moscow
117259, Russia.}}
\vskip 0.5 cm
\centerline{and}
\vskip 0.5cm
\centerline{J.J.M. Verbaarschot}
\vskip 0.2cm
\centerline{Department of Physics}
\centerline{SUNY, Stony Brook, New York 11794}
\vskip 1cm
\centerline{\bf Abstract}

Based on the chiral symmetry breaking pattern and the corresponding low-energy
effective lagrangian, we determine the fermion mass dependence of the
partition function and derive sum rules for eigenvalues of the QCD Dirac
operator in finite Euclidean volume. Results are given for $N_c  = 2$
and for Yang-Mills theory coupled to several light
adjoint Majorana fermions. They
coincide with those derived earlier in the framework of random matrix theory.

\vfill
\noindent
\begin{flushleft}
April 1994
\end{flushleft}
\eject
\pagestyle{plain}

\vskip 1.5cm
\renewcommand{\theequation}{1.\arabic{equation}}
\setcounter{equation}{0}
\noindent
\vskip 0.5cm

\section{Introduction.}

More than 20 years have passed since the advent of $QCD$, but the
crucial physical questions about the mechanism of confinement and
the structure of the $QCD$ vacuum state still remain basically unanswered.
The $QCD$ sum rule method \cite{Shif} proved to be very useful in
understanding the hadron spectrum and dynamics, but only
rather crude characteristics of the vacuum, such as the gluon
condensate $<G_{\mu\nu}^a G_{\mu\nu}^a>_0$ and quark condensate
$<\bar{q} q>_0$, could be determined in this way.

There are two other directions of research which seem now most promising:
lattice numerical simulations \cite{lat} and model simulations
\cite{Shur} which can be used to confront assumptions
on the form of the $QCD$ vacuum functional with experiment.
Naturally, both lattice and model simulations are performed in a finite
volume.

Recently, it has been observed that, besides experimental data, there are
some {\em exact} theoretical results specific for finite volume systems
which follow from first principles {\em and} the assumption that
chiral symmetry breaking occurs \cite{LS}. In particular, the light quark
mass dependence of the finite volume $QCD$ partition function has been
determined and, on basis of that, sum rules for the eigenvalues of
Euclidean Dirac operator have been derived. The simplest such sum rule is
\be
\label{sr3}
\left< \sum_n \frac 1{\lambda_n^2} \right>_\nu ~~=~~ \frac{(\Sigma V)^2}
{4(|\nu| + N_f)},
  \ee
where $V$ is the 4-dimensional volume, $\Sigma = |<\bar{\psi} \psi>_0|$,
and $\nu = (1/32\pi^2) \int d^4x G_{\mu \nu}^a {\tilde{G}}_{\mu \nu}^a$ is the
topological charge of the gauge field configuration.
The averaging goes over all gauge fields with given $\nu$ with the weight
\[ \propto \exp \left\{- \frac 1{4g^2} \int d^4x G_{\mu \nu}^a G_{\mu \nu}^a
\right \} \left[ {\det}'i{\cal D} \right]^{N_f},   \]
where ${\cal D}$ is the massless Dirac operator and $\det'$ is the product of
all its {\em nonzero} eigenvalues (the index theorem enforces the
appearance of $\nu$ zero fermion modes for each flavor in the gluon
background with topological charge $\nu$. They are shifted from zero when a
small fermion mass $m$ is switched on).
The sum in LHS of (\ref{sr3}) runs only over
{\em positive} eigenvalues $\lambda_n$ which are
{\em much less} than the characteristic hadron scale $\propto \Lambda_{QCD}$
(The region $\lambda_n \gg \Lambda_{QCD}$ brings about the ultraviolet
divergent contribution $\propto \Lambda_{ultr}^2 V$ which bears no
interesting dynamic information). The result (\ref{sr3})
is valid when the length of the box $L$ is much larger than
$\Lambda_{QCD}^{-1}$.

The sum rule (\ref{sr3}) may be checked by lattice and model simulations, or
more correctly, be used as a test for the correctness of these
simulations. Indeed, if we approximate the sum over all gauge configurations
by a liquid of instantons, the sum rule (1.1) is
satisfied \cite{SHURYAK-VERBAARSCHOT-1993}.

The sum rule (\ref{sr3}) has been derived under the
assumption of the standard pattern of chiral symmetry breaking
  \be
  \label{cb3}
SU_L(N_f) \otimes SU_R(N_f) \rightarrow SU_V(N_f).
  \ee
However, the breaking according to this scheme only occurs for fermions
belonging  to the complex fundamental representation of the color group. This
is the case when $N_c \geq 3$.
For $N_c = 2$, the fundamental representation is pseudoreal: quarks and
antiquarks transform in the same way under the action of the gauge group
and the pattern of chiral symmetry breaking {\em is} different leading
to different sum rules \cite{JJV-1994}. A third pattern of chiral symmetry
breaking is for fermions in the adjoint (real) representation leading to yet
another class of sum rules \cite{LS,JJV-1994}.

As we
shall discuss in more detail in the next section, the true chiral symmetry
group of the lagrangian of $SU(2)-$color with fundamental fermions
is $SU(2N_f)$ rather than just $SU_L(N_f) \otimes SU_R(N_f)$
(it involves also the transformations that mix
quarks with antiquarks). The pattern of the spontaneous symmetry breaking
due to formation of a quark condensate is
  \be
  \label{cb2}
    SU(2N_f) \rightarrow Sp(2N_f).
  \ee

One of the main results of this paper is the derivation of the analog
of sum rule (\ref{sr3}) for the case $N_c = 2$:
   \be
  \label{sr2}
\left< \sum_n \frac 1{\lambda_n^2} \right>_\nu ~~=~~ \frac{(\Sigma V)^2}
{4(|\nu| + 2N_f-1)}.
  \ee
It is of significant practical importance as numerical simulations of
$QCD$ with $N_c = 2$ are much easier and the sum rule (\ref{sr2}) may be
checked sooner.

In \cite{LS}, the case of Majorana fermions belonging to adjoint
representation of the color group has been also considered. The pattern of
symmetry breaking is
   \be
  \label{cbad}
    SU(N_f) \rightarrow SO(N_f).
  \ee
The analog of the sum rule (\ref{sr3}) was derived in \cite{LS}
for $N_f = 1$ and $N_f = 2$ which are technically simpler. In this paper we
fill up this gap and derive the sum rule for arbitrary $N_f$:
   \be
  \label{srad}
\left< \sum_n \frac 1{\lambda_n^2} \right>_{\bar{\nu}}
 ~~=~~ \frac{(\Sigma V)^2}{4(|\bar{\nu}| + (N_f+1)/2)},
  \ee
where $\bar{\nu} = \nu N_c$. The sum runs over the positive doubly
degenerate \cite{LS} pairs of eigenvalues of the adjoint Dirac operator.

Actually, the results (\ref{sr2}) and (\ref{srad}) are not new. They have
been obtained earlier using universality arguments and random matrix theory
\cite{JJV-1994}. In that approach it was argued that the spectrum of the Dirac
operator near zero virtuality is universal, $i.e.$ it is completely determined
by the symmetries of the system and can therefore be obtained by a random
matrix theory with only these symmetries as input.
The aim of this
paper is to illustrate once more that both approaches (stochastic matrices
and chiral lagrangian) are physically equivalent. The results (\ref{sr3}),
(\ref{sr2}) and (\ref{srad}) follow
from only the symmetry properties of the theory
which are properly accounted for in both ways of reasoning.

The structure of the paper is the following. In the next section, we
discuss symmetry breaking patterns for $QCD$ with $N_c=2$ and for the
theory with adjoint fermions. In section 3, we discuss
the technically simpler case $N_c = N_f = 2$,
 where a nice closed expression for
the finite volume partition function can be derived. In section 4, we derive
sum rules (\ref{sr2}) and (\ref{srad}) for any $N_f$ and any topological charge
sector. In section 5, we
derive expressions for partition function in the sector with a given
topological charge for the case of equal fermion masses both for $N_c = 2$ with
fundamental fermions and for
adjoint theories. The results, which generalize the
corresponding expression for $QCD$ with $N_c \geq 3$ derived in
\cite{LS},  can be expressed in terms of a determinant of antisymmetric
matrices made of Bessel functions (fundamental fermions with $N_c = 2$) or
certain integrals related to series of Bessel functions (adjoint fermions).
\vskip 1.5cm

\section{Chiral symmetry and its breaking.}
\renewcommand{\theequation}{2.\arabic{equation}}
\setcounter{equation}{0}

In this section we discuss the chiral symmetry breaking pattern for $SU(2)$
color with fundamental fermions and for adjoint fermions with $SU(N_c), \,
N_c\ge 2$. The discussion up to (2.9) will be in Minkowsky space. All
other sections of this paper and the remainder of this section starting from
(2.10) will be in Euclidean space time.

  The fermion part of the QCD lagrangian involving $N_f$ massless quark
Dirac fields belonging to the fundamental representation of the color group
is habitually written as
  \be
  \label{Lferm}
{\cal L}_{ferm} ~=~i \sum_{f=1}^{N_f} \left[ \bar{q}^f_L {\cal D} q^f_L +
 \bar{q}^f_R {\cal D} q^f_R \right],
  \ee
where $q_{L,R} = \frac 12 (1 \pm \gamma^5)q$ and ${\cal D}$ is the Dirac
operator. In this basis the  $U(N_f) \otimes U(N_f)$
symmetry of the lagrangian is immediately obvious.
Of course, a $U_A(1)$ subgroup
is broken explicitly  by the anomaly.

However, for $N_c = 2$, the symmetry is higher. To see
that, one should note that the combination
  \be
  \label{qprime}
\tilde{q}^f_{iL} ~~=~~\epsilon_{ij} C \bar{q}^{jf}_R
  \ee
is a left-handed spinor that transforms according to the same
representation of the $SU(2)$ color group as $q^f_{iL}$ ($i$ is the color
index and $C$ is the charge conjugation matrix). It is convenient to write
(\ref{Lferm}) in terms of $N_f + N_f$ two-component Weyl spinors $w_\alpha^f$
    \be
  \label{LWeyl}
{\cal L}_{ferm} ~=~i \sum_{f=1}^{2N_f} \bar{w}^f {\cal D} w^f.
  \ee
Obviously, this lagrangian enjoys the $U(2N_f)$ symmetry. The $U(1)$ part of
it is anomalous and only the $SU(2N_f)$ symmetry is left in the full quantum
theory. A subgroup is broken spontaneously, however, by the formation of
the fermion condensate. The condensates with maximal flavor symmetry
\cite{PESKIN-1980} are given by
  \be
  \label{cond2}
\left< \epsilon^{ij} \epsilon^{\alpha \beta} w^f_{i\alpha} w^{f'}_{j\beta}
\right> _0 ~~=~~ \frac {\Sigma}2 I^{ff'}.
  \ee
The matrix $I$ is the  $2N_f \times 2N_f$ antisymmetric matrix
\be
  \label{I}
I = \left (
\begin{array}{cc} 0 & {\bf 1} \\
                 -{\bf 1} & 0
\end{array}
\right ),
\ee
where {\bf 1} is the $N_f \times N_f$ unit matrix.
The fact that $I^{ff'}$ is antisymmetric follows just from the Grassmann
nature of the fields $w^f_{i\alpha}$. The specific form (\ref{I}) (It is
fixed up to an arbitrary transformation $I \rightarrow UIU^T$ with $U \in
SU(2N_f)$ ,  $U^T$ is the transpose of $U$. Multiplying of the condensate
matrix by an overall $U(1)$ phase factor
is also possible but it corresponds to going into a sector with a
different vacuum angle $\theta$.) follows from the requirement that the
formation of the condensate $does$ $not$
lead to the spontaneous breaking of the vector  symmetry $SU_V(N_f)$.
This is a consequence of the Vafa-Witten theorem \cite{VW}:
the spontaneous breaking of a vector flavor symmetry would result in the
appearance of scalar Goldstone particles. However, the vector-like nature of
$QCD$ enforces the lightest particle in the spectrum to be
a pseudoscalar
\footnote{Earlier, only the
case $N_c \geq 3$ when the
fermions belong to the essentially complex representation has been
considered, but the
theorem can be easily generalized to the case when the representation is
real or pseudoreal \cite{Shif1}.}
\cite{Weingarten-1983,Witten-1983,Nussinov}.
The same conclusion follows from an analysis of the effective potential
for $N_c \to \infty$ \cite{Coleman-Witten-1980}.

Naturally, the mere fact that the condensate is formed and that the symmetry is
broken cannot be proven rigorously and {\em is} an assumption. For standard
$QCD$ with $N_c = 3$ , the breaking of chiral symmetry is an experimental
fact. It is natural to think that a similar (not yet quite disclosed) mechanism
also leads to the formation of a chiral condensate in the other theories
discussed above.
The transformations of $SU(2N_f)$ which leave the form $I^{ff'} w_f w_{f'}$
invariant constitute the symplectic group $Sp(2N_f)$ (of which $SU_V(N_f)$ is
a subgroup), and we arrive at the chiral breaking pattern (\ref{cb2}).

This result is known in the literature. People were interested with
nonstandard patterns of chiral breaking mainly in association with
technicolor models \cite{PESKIN-1980,DPKSV}. In a recent work \cite{DP}, this
question has also been addressed, but the  authors assumed a flavor asymmetric
condensate formation with a different symmetry breaking pattern.

The breaking (\ref{cb2}) leads to the appearance of
\[ (2N_f)^2 - 1 - (2N_f^2 + N_f) ~~=~~ 2N_f^2 - N_f -1  \]
Goldstone bosons which are parameterized by
the coset $SU(2N_f)/Sp(2N_f)$. For $N_f =2$,
we have 5 instead of the usual 3 Goldstone bosons.

Up to now, we have considered only massless fermions. The mass term
  \be
  \label{Lm}
  {\cal L}_m ~~=~~\epsilon^{ij} \epsilon^{\alpha \beta} {\cal M}_{ff'}
w^f_{i\alpha} w^{f'}_{j\beta} ~+~ c.c.
  \ee
(where ${\cal M}_{ff'}$ is an arbitrary antisymmetric complex matrix $2N_f
\times 2N_f$) breaks the $SU(2N_f)$ symmetry explicitly and gives masses to the
Goldstone bosons which are,
however, small if the matrix elements ${\cal M}_{ff'}$
are much smaller than $\Lambda^{QCD}$ which we will always assume.

Let us discuss now the symmetry breaking pattern in the case when the
fermions belong to the adjoint representation of the color group which is
real \cite{PESKIN-1980,DPKSV,LS}. The fermion part of the
lagrangian\footnote{
The lagrangian (\ref{Lfermad}) can also be expressed in terms of the
Majorana 4-component fields
\[
\lambda_M = \left(
\begin{array}{c} w\\ -\sigma_2\bar w   \end{array} \right). \]
(see e.g. \cite{Ramond}).}.
is
    \be
  \label{Lfermad}
{\cal L}^{ad}_{ferm} ~=~i \sum_{f=1}^{N_f} \bar{w}^f {\cal D} w^f,
  \ee
where $N_f$ is now the number of massless Weyl adjoint fermions $w^{fa}$
The lagrangian (\ref{Lfermad})
possesses a $U(N_f)$
symmetry of which the $U(1)$ part is anomalous. In this case the Vafa-Witten
theorem dictates \cite {Shif1} the condensate to be {\em
flavor-symmetric} (see also \cite{PESKIN-1980}),
  \be
  \label{condad}
\left< \epsilon^{\alpha \beta} w^{af}_{\alpha} w^{af'}_{\beta}
\right > _0 ~~=~~ \frac{\Sigma}2 \delta^{ff'},
  \ee
(This form is fixed up to an arbitrary unitary transformation from $SU(N_f)$
: $\delta^{ff'} \rightarrow (UU^T)^{ff'}$.)
The formation of the condensate breaks down the symmetry
to $SO(N_f)$ (Only orthogonal transformations leave the form
$\epsilon^{\alpha \beta} w^{af}_{\alpha} w^{af}_{\beta}$ invariant). This
leads to the appearance of
\[ N_f^2 - 1 - \frac {N_f(N_f-1)}2 ~=~ \frac {N_f(N_f+1)}2 ~-~1 \]
Goldstone bosons. They acquire small masses after switching on the quark
mass term
   \be
  \label{Lmad}
  {\cal L}_m^{ad} ~~=~~\epsilon^{\alpha \beta} {\cal M}_{ff'}
w^{af}_{\alpha} w^{af'}_{\beta} ~+~ c.c.,
  \ee
(where ${\cal M}_{ff'}$ is now a {\em symmetric} complex $N_f \times
N_f$) matrix.

Before proceeding further, let us discuss the simplest case $N_f = 1$ and
remind simultaneously some relevant arguments from \cite{LS} where the reader
can find the details. It is well known that, in standard $QCD$ with
only one light quark, no spontaneous chiral symmetry breaking occurs at
all. After explicit breaking of $U_A(1)$ due to the axial anomaly, the
symmetry of the
quantum lagrangian is just $U_V(1)$ which is not broken spontaneously by
the formation of a condensate.

Also in the other theories discussed above there is no
spontaneous symmetry breaking for one flavor.
For the adjoint $N_f = 1$ case (for zero fermion mass , this theory is
just $N=1$ supersymmetric Yang-Mills), no symmetry is left right from the
beginning. For $N_c=2$, $N_f=1$, the chiral symmetry is $SU(2)$ but it is
not broken by the formation of the condensate $<\epsilon^{ij}
\epsilon^{\alpha \beta} w_{i\alpha}w_{j\beta} >$ which is the group
invariant.

That means that the spectrum of all $N_f=1$ theories involves a gap and,
when the size of the system $L$ is much larger than the characteristic
scale $\Lambda_{QCD}^{-1}$, the extensive property for the partition
function holds:
  \be
  \label{extens}
Z(m,\theta) = \exp (-\epsilon_{vac}(m, \theta) V )~\sim ~ \exp (
\Sigma m \cos \theta V ),
  \ee
where $m \ll \Lambda_{QCD}$ is the quark mass and $\theta$ is the vacuum
angle. The normalization $\epsilon_{vac}(0,\theta) = 0$ is chosen. The
particular combination $m \cos \theta$ appears in the first term of the
Taylor expansion of $\epsilon_{vac}$ in $m$ due to  Ward identities which
enforce all physical quantities to depend on the parameters $m$ and
$\theta$ in the combination $me^{i\theta}$ and the requirement of reality
of $\epsilon_{vac}$. The parameter $\Sigma$ has the
meaning of the quark condensate. For definiteness, we have
written (\ref{extens}) for theories with fundamental fermions. For
adjoint fermions, the combination $me^{i\theta/N_c}$ enters and one should
substitute $\theta \rightarrow \theta/N_c$  in (\ref{extens}).

The partition function in the sector with a given topological charge $\nu$
is given by the Fourier integral
  \be
  \label{ZnuN1}
Z_\nu(m) ~=~ \frac 1{2\pi} \int_0^{2\pi} e^{-i\nu\theta} d\theta Z(m,\theta)
{}~=~I_\nu(m\Sigma V),
  \ee
where $I_\nu(x) = I_{-\nu}(x)$ is the exponentially rising modified Bessel
function. In
the adjoint case, the integral over $\theta$ extends up to $2\pi N_c$ and
the same result holds but with rescaling $\nu \rightarrow \bar{\nu} =
\nu N_c$. The admissible topological numbers $\nu$ are then integer multiples
of $1/N_c$ \cite{frac}.
The sum rules (\ref{sr3}), (\ref{sr2}), and (\ref{srad}) are
obtained from the expansion of $Z_\nu(m)$ in $m$ and
comparing it with the expansion of (the Euclidean gamma matrices are
anti-hermitean)
  \be
  \label{detrat}
\left< \frac {\det'(i{\cal D} - m)}{\det'(i{\cal D})} \right>_\nu ~=~
\left< \prod_{\lambda_n > 0} \left(1 + \frac {m^2}{\lambda_n^2} \right)
\right>_\nu.
  \ee
They all {\em coincide} for
$N_f =1$.
\vskip 1.5cm
\section{$N_c = N_f = 2.$}
\renewcommand{\theequation}{3.\arabic{equation}}
\setcounter{equation}{0}
 When $N_f \geq 2$, spontaneous chiral symmetry breaking occurs and
Goldstone modes appear. The presence of quasi-massless particles in the
spectrum
brings about long-distance correlations which do not allow one to write down
the finite volume partition function in the extensive form (\ref{extens}).
However, the properties of Goldstone bosons are known,
and one can take them into
account explicitly in the path integral for the partition function. When
$L$ is much less than the inverse Goldstone mass $\sim (m\Lambda_{QCD})^{-1
/2}$, it suffices to take into account only the zero spatial Fourier harmonics
of the Goldstone fields and the path integral is reduced to a finite
dimensional integral over the coset space where the Goldstone modes live.

The case $N_f = 2$ is particularly simple because of the simple structure of
these cosets. The integrals can be done explicitly. For $N_c =3$ with
fundamental fermions,
the Goldstone modes live on $SU(2) \equiv S^3$. In the adjoint
case, the coset is $SU(2)/SO(2) = S^2$. These two cases have been treated
in \cite{LS}. The analysis of the $N_c = N_f = 2$ theory is performed
along the same lines.

Again, the coset is simple:
  \be
  \label{coset}
{\cal K} ~=~SU(4)/Sp(4) ~\equiv~ SO(6)/SO(5) ~=~ S^5.
  \ee
The corresponding low-energy chiral effective lagrangian belongs to the
class considered in \cite{HL}. It can be formulated in terms of the
6-dimensional vector {\bf n} with the constraint ${\bf n}^2 = 1$. When
$L \ll (m\Lambda_{QCD})^{-1/2}$, the kinetic term
$\sim (\partial_\mu {\bf n})^2$ in the effective lagrangian can be
neglected, and only the mass term is relevant. If all quark masses are equal
and real, the mass term is just $2m\cos \frac \theta 2 \Sigma\, n_6 \, $. As
earlier, the particular dependence on the vacuum angle $\theta$ follows
from the Ward identities. The partition function can be written as
  \be
  \label{intS5}
Z(m,\theta) \propto \int d\Omega_5 e^{2m\cos \frac \theta 2 \Sigma V
\cos \Phi},
  \ee
where $\Phi$ is the polar angle on $S^5$ (so that $d\Omega_5 \propto \sin^4
\Phi$). The integral is elementary and we get
  \be
  \label{ZNf2}
Z(m, \theta) ~=~ \frac {2I_2(2m\cos \frac \theta 2 \Sigma V)}
{(m\cos \frac \theta 2 \Sigma V)^2},
  \ee
(we choose the normalization $Z(0, \theta) = 1$). The partition function in
a given topological sector $\nu$ is the coefficient of $e^{i\nu\theta}$ in
the Fourier expansion of $Z(m, \theta)$:
  \be
  \label{ZnuNf2}
Z_\nu ~=~ 2  \sum_{k=0}^\infty \left( \frac x2 \right)^{2(|{\nu} | +k)}
\frac {[2(|{\nu} | +k)]!}{k! (|{\nu} | + k)! (|{\nu} | + k +2)!
(2|{\nu} | + k)!}
  \ee
with $x = m\Sigma V$.
 Comparing the coefficients of $x^{|{\nu}|}$ and $x^{|{\nu}|+
2} $, we arrive at the sum rule (\ref{sr2}) with $N_f = 2$.

A more explicit expression for the partition function $Z_{\nu}$ is obtained
is we integrate over $\theta$ first. As a result we find
\be
\label{intI2nu}
Z_\nu = \frac 8{3\pi}\int_0^\pi d\Phi \sin^4 \Phi I_{2\nu}(2x\cos\Phi).
\ee
By writing $\sin^4 \Phi  = \frac 18(\cos 4\Phi - 4\cos 2\Phi +3)$, this
integral can be expressed into known integrals \cite{GRADSHTEYN-RHYZIK}. The
result is
\be
\label{pfaf2}
Z_\nu = \frac 13 \left[ 3I_{\nu}^2(x) - 4 I_{\nu -1}(x) I_{\nu +1}(x) +
I_{\nu -2}(x) I_{\nu +2}(x) \right].
\ee
In section 5 we will show that this formula can be generalized to arbitrary
$N_f$ in the form of the Pfaffian of an anti-symmetric matrix to be defined
later.

\vskip 1.5cm
\renewcommand{\theequation}{4.\arabic{equation}}
\setcounter{equation}{0}
\section{\bf  Sum rules for any $N_f$.}
Let us consider first the case of adjoint fermions with arbitrary number of
flavors $N_f$. The coset $SU(N_f)/SO(N_f)$ is spanned by the symmetric
unimodular matrices $S$ which can be presented in the form $S = UU^T$ with
$U \in SU(N_f)$ . The partition function
$Z({\cal M},\theta)$ represents the invariant integral over the coset which,
for convenience, will be traded
off for the integral over the full group $SU(N_f)$:
  \be
  \label{ZMthet}
Z({\cal M}, \theta) = \int_{SU(N_f)} dU \exp\{V\Sigma Re [Tr({\cal M}
e^{i\theta /N_c N_f} UU^T)] \}.
  \ee
Here, $dU$ is the Haar measure on the group, and due to Ward identities,
the mass matrix ${\cal M}$
and the vacuum angle $\theta$ enter in a particular combination ${\cal M}
\exp(i\theta / N_c N_f)$ only \cite{LS}. The integral (\ref{ZMthet})
involves also dummy variables spanning the stability subgroup of orthogonal
$SO(N_f)$ matrices on which the integrand does not depend.

The partition function $Z_\nu({\cal M})$ in the sector with a given
topological charge $\nu$ is given by the Fourier integral
 \be
 \label{ZnuM}
Z_{\nu}({\cal M}) ~=~ \frac 1{2\pi N_c} \int_0^{2\pi N_c}
e^{-i\nu\theta} d\theta Z({\cal M},\theta).
 \ee
It can be written as the integral over the full $U(N_f)$ group as
$(\bar\nu = \nu N_c)$
  \be
  \label{ZnuX}
Z_{\bar \nu}(X) = \int_{U(N_f)} dU \left
(\det U  \right )^{-2\bar{\nu}} \exp\left ({\rm Re}
 ({\rm Tr} X U U^T) \right ),
\ee
where $X \equiv V\Sigma {\cal M}$.
 The mass matrix ${\cal M}$
is symmetric and complex.

The partition function (\ref{ZnuX}) is an invariant function of $X$:
\be
\label{inv}
Z_{\bar\nu}(V^TXV ) =(\det V)^{2\bar{\nu}} Z_{\bar\nu}(  X ),
\ee
where $V$ is a unitary matrix. This enables us to expand the partition
function for $\bar\nu \ge 0$ as
\be
 \label{expan}
Z_{\bar\nu}(X) = N_{\bar\nu} [\det(X)]^{\bar{\nu}}
( 1 + a_{\bar\nu} {\rm tr} X^\dagger X + {\cal O}(X^4)).
\ee
The normalization factor is denoted by $N_{\bar\nu}$.
(For $\bar\nu < 0$ the first factor in the expansion should be replaced
by $[\det(X^\dagger)]^{-\bar{\nu}}$.)
Such integrals have been considered before in \cite{LS}
for the $SU_L(N_f)\otimes SU_R(N_f) \rightarrow SU_V(N_f)$ chiral symmetry
breaking scheme.
A similar analysis will be applied to the present partition function. The
matrix $X$ can be decomposed as
\be
 \label{Xs}
X= \sum_{a=1}^{M^s} x_a t^a,
\ee
where $t^a$ is an orthonormal set of symmetric real matrices
which are the generators for the symmetric unitary matrices.
They satisfy the identities
\be
 \label{norm}
{\rm Tr} \{t^a t^b\} &=& \frac 12 \delta^{ab},\qquad  a = 1, \cdots, M^s,
\nonumber\\
\sum_a t^a t^a &=& \frac {M^s}{2 N_f},
\ee
The total number of generators $M^s = N_f(N_f +1)/2$. The coefficients
$x_a$ in (\ref{Xs}) are complex numbers. For any symmetric
matrices $A$ and $B$, we have the identity
\be
 \label{set}
\sum_{a=1}^{M^s} {\rm Tr}(t^a A){\rm Tr}(t^a B) = \frac 12 {\rm Tr}(AB).
\ee
This enables us to show that the partition function (\ref{ZnuX}) satisfies
the differential equation
\be
 \label{diff}
\sum_a \del_{x_a}\del_{\bar x_a} Z_{\bar\nu} = \frac{N_f}{8} Z_{\bar\nu}.
\ee
 Substituting here the expansion (\ref{expan}), the coefficients $a_{\bar\nu}$
can be easily found:
\be
a_{\bar\nu} = \frac 1{4 \left(|\bar{\nu}| + ({N_f + 1})/2 \right)}.
\ee
Comparing (\ref{expan}) with the mass expansion of the averaged Dirac
determinant \cite{LS}, one immediately gets to the spectral sum rule
(\ref{srad}) which agrees with the result obtained by using
random matrix theory \cite{JJV-1994}.

The calculation of the sum rule for fermions in the fundamental $SU(2)-$color
representation is analogous. In this case the partition function
corresponds the chiral symmetry breaking scheme $SU(2N_f) \rightarrow
Sp(2N_f)$. The coset is spanned by the antisymmetric unimodular matrices
$A$ which can be presented in the form $A = UIU^T$ where $U \in SU(N_f)$ and
$I$ is the symplectic matrix defined in (\ref{I}). In this case the mass enters
only in the combination ${\cal M} \exp(i\theta/N_f)$.
The partition function in the sector with a given
topological charge $\nu$ is given by the integral over the coset and vacuum
angle $\theta$ which can be expressed in terms of the integral over the
full $U(2N_f)$ group :
\be
 \label{ZnuXa}
Z_\nu(X) = \int_{U(2N_f)} dU \left
(\det U  \right )^{-\nu} \exp\left (\frac 12 {\rm Re}
({\rm Tr}{ X} U I U^T) \right ).
\ee
The matrix $X = V\Sigma {\cal M}$ is now an anti-symmetric complex
matrix. This partition function also satisfies
an invariance relation
  \be
  \label{inv1}
Z_{\nu}(V^TXV ) =(\det V)^{\nu} Z_{\nu}(  X ),
  \ee
(cf. (\ref{inv})), and therefore can be expanded in group
invariants. The expansion is a little bit modified as compared to (\ref{expan})
and for $\nu \ge 0$ it has the form
  \be
  \label{exp1}
Z_\nu(X) = N_\nu [ {\rm Pf}(X)]^{{\nu}}
( 1 + \frac 12 a_\nu {\rm tr} X^\dagger X + {\cal O}(X^4)).
  \ee
where ${\rm Pf}(X) = \sqrt{\det( X)}$ is the Pfaffian of the
antisymmetric matrix $ X$. (For $\nu < 0$ the first factor in this
expansion has to be replaced by $[{\rm Pf}(X^\dagger)]^{{-\nu}}.$)

The matrices $UIU^T$ are anti-symmetric complex
unitary matrices. Their generators are anti-symmetric real $2N_f\times 2 N_f$
matrices which we use as a basis for the expansion of X:
\be
X= \sum_{a=1}^{M^{as}} x_a t^a.
\ee
In this case the total number of generators is $M^{as} = N_f(2N_f -1)$.
Instead of (\ref{norm}) we have the relations
\be
{\rm Tr} \{t^a t^b \}&=& \frac 12 \delta^{ab},\qquad  a = 1,
\cdots, M^{as},\nonumber\\ \sum_a t^a t^a &=& \frac {M^{as}}{4 N_f},
\ee
and  (\ref{set}) is also valid for anti-symmetric matrices $A$ and $B$. In this
case
the sum over $a$ runs over the anti-symmetric generators.
The partition function (\ref{ZnuXa}) satisfies the differential equation
\be
 \label{diff1}
\sum_a \del_{x_a}\del_{\bar x_a} Z_\nu = \frac{N_f}{16} Z_\nu.
\ee
(cf. (\ref{diff})).
Substitution of the expansion of $Z_\nu$ in powers of $X$ yields
\be
a_\nu = \frac 1{4(|\nu| + 2N_f -1)}.
\ee
The corresponding sum-rule has the form (\ref{sr2}) which agrees with
the result obtained from random matrix theory.

The sum rules (\ref{sr3}), (\ref{sr2}), (\ref{srad}) can be written
universally as
\be
 \label{sruniv}
\left< \sum_{\lambda_n > 0} \frac 1{\lambda_n^2} \right>_\nu
 = \frac {(\Sigma V)^2}
{4 \left [|\nu| + ({\rm dim (coset)} + 1)/N_f\right ]},
\ee
with the rescaling $\nu \rightarrow \bar\nu$ and counting in the sum only
one eigenvalue of each degenerate pair in the adjoint case.

\vskip 1.5cm
\renewcommand{\theequation}{5.\arabic{equation}}
\setcounter{equation}{0}
\section{\bf  Partition function for equal masses}

In this section, we evaluate the partition function for the special case
when all masses are equal to $m$. We use the
notation $x = mV\Sigma$. First, we consider the case of adjoint fermions.
The matrix $ U U^T$ in (4.1) is a symmetric unitary
matrix which can be diagonalized by an orthogonal matrix. To evaluate
the integral we use the eigenvalues $\exp(i\theta_k)$
and the eigenvectors of $U U^T$ as new
integration variables. The Jacobian of this transformation has been
used extensively in random matrix theory. It has the form
  \be
  J_{s}(\theta_1, \cdots ,\theta_{N_f})  \propto
\prod_{k<l} |e^{i\theta_k} - e^{i\theta_l}|.
  \ee
The partition function can be rewritten as
\be
Z_{\bar{\nu}}(x) = \frac{\pi^{N_f/2}}{2^{N_f} \Gamma(1+N_f/2)}
\int_{-\pi}^{\pi}\cdots \int_{-\pi}^{\pi} \prod_{k=1}^{N_f}
\frac{d\theta_k}{2\pi} \prod_{k<l} |e^{i\theta_k} - e^{i\theta_l}|
\exp\left (x\sum_k \cos (\theta_k)
+i\bar\nu \sum_k \theta_k \right).\nonumber\\
  \ee
The normalization is such that $Z(0) = Z_0(0) = 1$.
Integrals of these type have been studied extensively by Mehta
\cite{MEHTA-1967,MEHTA-1991}.
The result can be expressed as a Pfaffian (see appendix {\bf A})
\be
\label{Zeven}
Z_\nu(x) = \frac{\pi^{N_f/2} N_f!}{2^{N_f} \Gamma(1+N_f/2)}
\,{\rm Pf}(A).
\ee
For even $N_f$, the matrix elements of
the anti-symmetric matrix $A$ are given
by
\footnote{Expanding $\epsilon(\theta - \phi)$ in the Fourier series on the
interval $-2\pi \leq \theta - \phi \leq 2\pi$, one can present the integral
 (\ref{Apq}) in the form
\[ A_{pq} = \frac 1\pi \sum_{k = -\infty}^\infty \frac 1{k + \frac 12}
 I_{\bar\nu + p + k + \frac 12 }(x) I_{\bar\nu  + q - k - \frac 12 }(x). \]}
\be
\label{Apq}
A_{pq} =
-i\int_{-\pi}^{\pi}\frac{d\theta}{2\pi}\int_{-\pi}^{\pi}\frac{d\phi}{2\pi}
\epsilon(\theta-\phi)\exp(i(p\phi+q\theta)) \exp(x\cos \phi + x\cos\theta +
i\bar\nu(\phi + \theta)).
\ee
The indices $p$ and $q$ run between $-\frac{N_f}2 + \frac 12 $ and
$\frac{N_f}{2} -\frac 12$.

For $\nu = 0$ and $x =0$, the Pfaffian in (5.3) can be calculated readily. The
result is
\be
\left .{\rm Pf}(A) \right |_{\nu = 0, \, x= 0} = \frac {2^{N_f/2}}
{\pi^{N_f/2} (N_f - 1)!!},
\ee
which gives the correct normalization for the partition function.

The simplest non-trivial case is $N_f = 2$ which describes two Majorana
 adjoint flavors (or
one Dirac flavor).
In this case, one can derive
\be
Z_{\bar\nu}(x) = \frac \pi 2 A_{-\frac 12~\frac 12}
= \frac {x^{2|\bar\nu|}}{(2|\bar\nu|+1)!}\left(1+\frac {x^2}{2|\bar\nu|+3}
 + {\cal O}(x^4) \right),
\ee
which agrees with results obtained in \cite{LS}.
For odd values of $N_f$ the partition function is equal to the Pfaffian
of the matrix $A$ supplemented by an additional row and column
(see appendix {\bf A})
\be
\label{Zodd}
Z_{\bar\nu} = \frac{\pi^{N_f/2}N_f!}{2^{N_f} \Gamma(1+N_f/2)}
{\rm Pf} \left (
\begin{array}{cc}   A &  c\\
                    -c^T & 0
\end{array} \right ),
\ee
where $c$ is a vector of length $N_f$ defined by
\be
c_p = (-1)^{(N_f-1)/2}\int_{-\pi}^\pi \frac{d\theta}{2\pi}
 \exp(ip\theta) \exp(x\cos\theta + i\bar\nu\theta).
\ee
Again, the Pfaffian can be calculated easily for $\nu =0$ and $x = 0$:
\be
\left .{\rm Pf}(A) \right |_{\nu = 0, \, x= 0} = \frac {2^{(N_f-1)/2}}
{\pi^{(N_f-1)/2} (N_f - 1)!!},
\ee
which results in a partition function that is normalized to one.

The small $x$ expansion of the partition functions (\ref{Zeven}), (\ref{Zodd})
starts from the term $\propto x^{N_f|\bar\nu|}$ which corresponds to the
presence of $N_f|\bar\nu|$ pairs of fermion zero modes in the gauge field
background with topological charge $\nu$. For $x \gg 1$,
\be
\label{asym}
Z_{\bar\nu} \propto \exp(N_f x)
\ee
for all $\bar\nu$ which means that, in the infinite volume limit, the
contributions of all topological charges are of the same order \cite{LS}.

For $N_c = 2$ with fundamental fermions, we proceed in exactly the same way.
In this case the anti-symmetric
unitary matrix $U I U^T$ can also be brought into the canonical form
by an orthogonal transformation. The result is
an anti-symmetric tri-diagonal
matrix with off-diagonal matrix elements $\pm \exp(i\theta_k)$.
Again we use the angles $\theta_k$ as
new integration variables. The Jacobian  of this transformation is
\be
J_{as}(\theta_1, \cdots ,\theta_{N_f})  \propto
\prod_{k<l} |e^{i\theta_k} - e^{i\theta_l}|^4,
\ee
leading to the partition function
\be
 Z_\nu = \frac 1{(2N_f - 1)!! N_f !} \int_{-\pi}^{\pi} \cdots \int_{-\pi}^{\pi}
\prod_{k=1}^{N_f} \frac {d\theta_k}{2\pi}
\prod_{k<l} |e^{i\theta_k} - e^{i\theta_l}|^4 \exp\left (x\sum_k \cos
(\theta_k)
+i\nu \sum_k \theta_k \right).\nonumber\\
\ee
The normalization factors have been chosen such that $Z_0(0) = 1$.
Also this partition function can be rewritten as a Pfaffian (see appendix {\bf
B}):
\be
Z_\nu = \frac 1{ (2N_f - 1)!!}{\rm Pf} (A),
\ee
where the matrix elements of $A$ are given by
\be
A_{pq} &=& (q-p)
\int_{-\pi}^{\pi} \frac{d\theta}{2\pi}  \exp(i(q+p)\theta)\exp(x\cos(\theta) +
i\nu\theta) \nonumber \\
 &=& (q-p) I_{p+q+\nu},
\ee
where the indices $p$ and $q$ run between $-(N_f-\frac 12)$ and $N_f -\frac
12$. Note that $A$ is a $2N_f \times 2 N_f$ dimensional matrix.

For $x \ll 1$, $Z_\nu(x) \propto x^{N_f|\nu|}$ which agrees with the zero
modes counting. The large $x$ asymptotics of $Z_\nu(x)$ is given by
(\ref{asym}) and is universal for all $\nu$.

As an example, let us consider $N_f =2$. In this case the matrix $A$
is given by
\be
A =
\left(
\begin{array}{cccc}
    0   &     I_{\nu -2}(x) & 2I_{\nu -1}(x) & 3I_{\nu}(x) \\
-I_{\nu -2}(x)  &   0     &   I_{\nu}(x)  &  2I_{\nu +1}(x) \\
-2I_{\nu -1}(x)  &  -I_{\nu}(x) &  0     &   I_{\nu +2}(x) \\
-3I_{\nu}(x)  &  -2I_{\nu +1}(x)  & -I_{\nu +2}(x) &  0    \\
\end{array}
\right ).
\ee
The Pfaffian of this matrix is given by
\be
{\rm Pf}(A) = 3I_{\nu}^2(x) - 4 I_{\nu -1}(x) I_{\nu +1}(x) +
I_{\nu -2}(x) I_{\nu +2}(x),
\ee
and $Z_\nu(x)$ coincides with our previous result (\ref{pfaf2})
derived by performing the integral over the angles
directly in (\ref{intI2nu}).
Expansion of $Z_\nu(x)$ in powers of $x$ reproduces the sum-rule derived in
previous sections.

\vskip 1.5cm
\renewcommand{\theequation}{6.\arabic{equation}}
\setcounter{equation}{0}
\section{ Discussion}
\vskip 0.5 cm
Depending on whether the representation of the color group is
real  ($ SU(N_c),\,\, N_c \ge 2$ with adjoint fermions)
, complex ($SU(N_c),\,\, N_c\ge 3$) or
pseudoreal ($SU(2)$ with fundamental fermions),
chiral symmetry breaking is realized
in a different way leading to different Goldstone sectors. They are
parameterized by $SU(N_f)/SO(2N_f)$, $SU_R(N_f)\times SU_L(N_f)/ SU(N_f)$
 and $SU(2N_f)/Sp(2N_f)$, respectively. For each of the three cases we have
determined explicitly the dependence of the partition function in finite
space-time
volume on the fermion mass $m$.  As was shown in
\cite{LS} , the comparison of the chiral expansion
of these partition functions with the chiral expansion of
the Dirac determinant in the original path integral
leads to sum rules for the inverse powers of
the eigenvalues of the massless Dirac operator in QCD. In other words, we have
very specific correlations between the eigenvalues of the Dirac operator.

The same triality is known from random matrix theory. The matrix elements
of the three classical random matrix ensembles are real, complex or
quaternion real.
These ensembles can be generalized to include the
chiral structure of QCD, and
it is possible to derive the spectral density and
its correlation functions. Previously it had been verified that for
complex representations of the color group,
this leads to exactly the same sum rules
as obtained from the finite volume partition function
\cite{VERBAARSCHOT-ZAHED-1993}. In the
framework of random matrix theory it is straightforward to derive sum rules for
the two other ensembles as well \cite{JJV-1994}. (The theories with
real, complex , and pseudoreal fermions
correspond to setting $\beta = 4$,  $\beta = 2$, and  $\beta = 1$ in the
universal formula (14) of Ref.\cite{JJV-1994}).

This leads to the question
whether the sum rules for the two remaining cases are indeed the ones
that follow from the effective partition
function that is based on general assumptions only. This question has been
answered affirmatively in this paper. Since both approaches are based
on symmetry this result did not come as a surprise. Apart from this, several
other new results were derived. In particular, we found explicit expressions
for the partition function for equal quark masses which generalize
results already known for the complex case.
This does not yet answer whether $all$ spectral correlations obtained
from random matrix theory are also properties of the QCD partition function.
However, from the study of spectra of classically chaotic systems, we know
that correlations on the order of one or several level spacings are universal
\cite{BOHIGAS-GIANNONI-1984}. It are precisely such correlations that are
responsible for the spectral sum rules. This leads to the conjecture that
the microscopic spectral density is universal.

In this paper, we always assumed that the parameters of the theory
were in the range
  \be
  \label{region}
m \ll \Lambda_{QCD},  \nonumber \\
\Lambda_{QCD}^{-1} \ll ~L~
\ll m_\pi^{-1} \sim (m\Lambda_{QCD})^{-1/2}.
   \ee
The upper bound for the length of the box $L$ was essential because,
if $L$ would be larger than the Compton wavelength of the Goldstone modes,
$\sim (m\Lambda_{QCD})^{-1/2}$, nonstatic modes of the Goldstone fields would
be
excited and the path integral could not be reduced to a finite dimensional
group integral. The calculation is also possible for larger $L$ by
exploiting the fact that the characteristic fluctuations of Goldstone
fields are small in that case \cite{GL}.
But, as we are mainly interested in the
derivation of sum rules (\ref{sr2}) and (\ref{srad}) which are formulated for
theories with {\em massless} fermions, results in the range
(\ref{region}) are quite sufficient for our purposes.

{}From the practical point of view, the results for the theory
with $N_c = N_f = 2$ analyzed in section 3 are
perhaps the most important.
This theory bears all features of standard $QCD$ but
involves fewer degrees of freedom so that it will be easier to confront
the sum rules with lattice simulations.

In this paper, the effective partition functions were postulated on the
basis of chiral symmetry and other general arguments. Our results
suggest that the static limit of the finite volume partition function
can also be derived directly from random matrix theory.
Work in this direction is under way.
\vskip 1.5cm
\noindent
{\bf Acknowledgements}
\vglue 0.4cm
 The reported work was partially supported by the US DOE grant
DE-FG-88ER40388. We are much indebted to A.I. Vainshtein
for many useful discussions and valuable suggestions and to M.A.Shifman for
pointing our attention to Refs.\cite{VW}-\cite{DPKSV}.
A.V.S. acknowledges a warm
hospitality during his stay at University of Minnesota.

\vskip 1.5cm
\renewcommand{\theequation}{A.\arabic{equation}}
\setcounter{equation}{0}
\noindent
\noindent{\bf\large Appendix A}
\vskip 0.5 cm
In this appendix we calculate integrals of the type
\be
\rho(u) = \int_{-\pi}^\pi \cdots \int_{-\pi}^\pi \prod_{k=1}^N u(\theta_k)
\frac{d\theta_k}{2\pi}
\prod_{k<l}^N |e^{i\theta_k} - e^{i\theta_l}|,
\ee
where $u(\theta)$ is an arbitrary function of $\theta$. We only give the main
steps and refer to Dyson's original paper \cite{DYSON-1962} or
the book of Mehta \cite{MEHTA-1967} for further details.
Because the integrand is a symmetric function of the integration variables, the
integration region can be restricted to the domain $D$ defined by
$-\pi\le \theta_1 \le \cdots
\le \theta_N\le \pi$ at the expense of a factor $N!$. For angles inside
this domain we have the identity
\be
\prod_{k<l} |e^{i\theta_k} - e^{i\theta_l}| = i^{-\frac{N(N-1)}2}
\det\{\phi_p(\theta_j)\},
\ee
where the determinant is over an $N\times N$ matrix with matrix elements given
by
\be
\phi_p(\theta_j) = \exp( ip\theta_j),
\quad p = -\frac{N-1}2, -\frac{N-3}2, \cdots, \frac{N-1}2.
\ee
To proceed we use the method of integration over alternate variables originally
due to \cite{MEHTA-1962}. The result for even $N$ is
\be
\rho(u) = i^{-\frac{N(N-1)}2} N!\int_{D} \prod_{k\,\,{\rm even}}
 \frac{d\theta_k}{2\pi} u(\theta_k)
\left |\begin{array}{ccccc}
 S_{\frac{1-N}2}(\theta_2) & \phi_{\frac{1-N}2}(\theta_2) & \cdots &
S_{\frac{1-N}2}(\theta_N) & \phi_{\frac{1-N}2}(\theta_N)\\
     .            &&&&   .    \\
     .            &&&&   .    \\
     .            &&&&   .    \\
 S_{\frac{N-1}2}(\theta_2) & \phi_{\frac{N-1}2}(\theta_2) & \cdots &
S_{\frac{N-1}2}(\theta_N)
& \phi_{\frac{N-1}2}(\theta_N)
\end{array}
\right |.\nonumber\\
\ee
The quantities $S_p(\theta_k)$ are defined as
\be
S_p(\theta_k) = \int_{-\pi}^{\theta_k} u(\phi) \phi_p(\phi) \frac{d\phi}{2\pi}.
\ee
For odd $N$, the last column is $S_{\frac{1-N}2}(\pi), \cdots,
S_{\frac{N-1}2}(\pi)$. The integrand
in (A.4) is a symmetric function of the integration variables, which permits
us to extend the integration range of each variable to $[-\pi, \pi]$ and
use the
integration theorem of appendix (A.7) of ref. \cite{MEHTA-1967}. As a result
we obtain
\be
\rho(u) = N! {\,\rm Pf} (A),
\ee
where the matrix elements of the anti-symmetric matrix $A$ are given by
\be
A_{pq} &=&-i \int_{-\pi}^\pi \frac{d\theta}{2\pi}[S_p(\theta) \phi_q(\theta) -
S_q(\theta) \phi_p(\theta)] \nonumber \\
          &=& -\frac i2 \int_{-\pi}^\pi\int_{-\pi}^\pi \frac{d\phi}{2\pi}
\frac{d\theta}{2\pi}
 \epsilon(\theta-\phi)u(\theta) u(\phi)[\phi_p(\phi) \phi_q(\theta)
-\phi_q(\phi) \phi_p(\theta)].
\ee
(the identity $i^{-\frac {N(N-1)}2} = (-i)^{\frac N2}$, even $N$, has been
used). The Pfaffian of an $N\times N$ ($N$ even)
anti-symmetric matrix is defined as
\be
{\rm Pf} (A) = \frac1{(N/2)!}
\sum_{\sigma} {\rm sign}(\sigma) A_{\sigma(1)\sigma(2)}
\cdots A_{\sigma(N-1)\sigma(N)},
\ee
where the sum is over all permutations $\sigma$ of $\{1,2,\cdots N\}$.
For Pfaffians we also have a Laplace expansion (see p. 393 of \cite{MUIR-1933})
\be
{\rm Pf} (A) = (-1)^{r+1} [A_{r1} {\rm Pf}(A_{1r,1r}) -
A_{r2} {\rm Pf}(A_{2r,2r}) - \cdots +(-1)^{N+1}A_{rN} {\rm Pf}(A_{Nr,Nr})],
\label{laplace}
\ee
where the matrix $A_{kr,kr}$ is obtained from matrix $A$ by deleting the
both the $k$'th and the $r$'th rows and columns.

For odd $N$, we expand the determinant with respect to the last column. Then
the theorem from appendix (A.7) of \cite{MEHTA-1967} can be applied
to each of the minors after extending the integration domain of each of the
integration variables to $[-\pi,\pi]$. The extra factor $1/[(N-1)/2]!$ cancels
against the same factor from the integration theorem in \cite{MEHTA-1967}.
Using the inverse of the Laplace expansion (\ref{laplace}) we find (with the
use of the identity $i^{-\frac {N(N-1)}2} = (i)^{\frac {N-1}2}$, odd $N$)
\be
\rho(u)
= N! {\rm Pf}\left (
\begin{array}{cc}   A &  c\\
                    -c^T & 0
\end{array} \right ),
\ee
where the matrix elements of $A$ are as defined above and
\be
 c_p = (-1)^{\frac {N-1}2}S_p(\pi).
\ee
\vskip 1.5cm
\renewcommand{\theequation}{B.\arabic{equation}}
\setcounter{equation}{0}
\noindent
\noindent{\bf\large Appendix B}
\vskip 0.5 cm
In this appendix we calculate the integral
\be
\rho(u) = \int_{-\pi}^\pi \cdots \int_{-\pi}^\pi \prod_k^{N}
 u(\theta_k)\frac{ d\theta_k}{2\pi}
\prod_{k<l}^N |e^{i\theta_k} - e^{i\theta_l}|^4.
\ee
According to Mehta, the Jacobian can be rewritten as a determinant of a
$2N \times 2N$ matrix:
\be
\prod_{k<l}^N |e^{i\theta_k} - e^{i\theta_l}|^4 = \det\{ \chi_p(\theta_j),
p\chi_p(\theta_j)\},
\ee
where
\be
\chi_p(\theta_j) = e^{ ip\theta_j},\quad p = -N+\frac 12, \cdots,
N-\frac 12.
\ee
In (B.2), the rows of the determinant are indexed by $p$, whereas the columns
are indexed by $j$. According to the theorem in appendix (A.7) of
\cite{MEHTA-1967} the integral (B.1) can now be expressed as a Pfaffian
\be
\rho(u) = N! \,{\rm Pf} (A),
\ee
where the matrix elements of $A$ are given by
\be
A_{pq} = \int_{-\pi}^\pi \frac{d\theta}{2\pi}
 u(\theta)\left (\chi_p(\theta)[q\chi_q(\theta)]
-\chi_q(\theta)[p\chi_p(\theta)]\right ).
\ee
\vglue 0.6cm

\vfill
\eject
\newpage
\setlength{\baselineskip}{15pt}

\bibliographystyle{aip}

\begin{thebibliography}{10}

\bibitem{Shif} Reprint volume {\it 'Vacuum Structure and QCD Sum Rules'}
(M.A. Shifman, ed.), North-Holland, 1992.

\bibitem{lat}
{\it Lattice 1992: Proceedings of the international symposion on
lattice field theory}, (J. Smit and P. van Baal eds.),
North-Holland, Amsterdam 1993.

\bibitem{Shur}
D.I.~Diakonov and V.Yu.~Petrov, Nucl. Phys. {\bf B245} (1984) 259;
E. Shuryak and J. Verbaarschot, Nucl. Phys. {\bf B341} (1990) 1;
Nucl. Phys. {\bf B 410} (1993) 55.

\bibitem{LS}
H.~Leutwyler and A.~Smilga,
\newblock Phys. Rev. {\bf D46} (1992) 5607.

\bibitem{SHURYAK-VERBAARSCHOT-1993}
E.~Shuryak and J.~Verbaarschot,
\newblock Nucl. Phys. {\bf A560} (1993) (306).

\bibitem{JJV-1994}
J. Verbaarschot, {\it The  spectrum of the Dirac operator and chiral
random matrix theory: the threefold way}, Phys. Rev. Lett. (1994)
(in press).

\bibitem{PESKIN-1980}
M. Peskin, Nucl. Phys. {\bf B175} (1980) 197.

\bibitem{VW}
C. Vafa and E. Witten, Nucl. Phys. {\bf B 234} (1984) 173.

\bibitem{Weingarten-1983}
D. Weingarten, Phys. Rev. Lett. {\bf 51} (1983) 1830.

\bibitem{Witten-1983}
E. Witten, Phys. Rev. Lett. {\bf 51} (1983) 2351.

\bibitem{Nussinov}
S. Nussinov and M. Spiegelglas, Phys. Lett. {\bf B181} (1986) 134.


\bibitem{Coleman-Witten-1980}
S. Coleman and E. Witten, Phys. Rev. Lett. {\bf 45} (1980) 100.

\bibitem{Shif1}
M. Shifman, {\it Private communication.}

\bibitem{DPKSV}
S. Dimopoulos, Nucl. Phys. {\bf B168} (1980) 69;
M. Vysotskii, I. Kogan and M. Shifman,
Sov. J. Nucl. Phys. {\bf 42} (1985) 318.


\bibitem{DP}
D.I. Diakonov and V.Yu. Petrov,
\newblock in '{\it Quark Cluster Dynamics}', Proceeding of the 99th WE-Heraeus
Seminar, Bad Honnef, 1992 (eds. K. Goeke $et$ $al.$), Springer 1993.

\bibitem{Ramond}
P. Ramond, {\it Field Theory: A Modern Primer}, Benjamin/Cunnings, Reading,
MA, 1981.

\bibitem{frac}
E. Cohen and C. Gomez, Phys. Rev. Lett. {\bf 52} (1984) 237; A. Zhitnitsky,
Nucl. Phys. {\bf B340} (1990) 56.

\bibitem{HL}
P. Hasenfratz and H. Leutwyler, Nucl. Phys. {\bf B343} (1990) 241.

\bibitem{BOHIGAS-GIANNONI-1984}
O.~Bohigas, M.~Giannoni, \newblock
in '{\it Mathematical and computational methods in
nuclear physics}', J.S. Dehesa et al. (eds.), Lecture notes in Physics
{\bf 209}, Springer Verlag 1984, p. 1;
O.~Bohigas, M.~Giannoni, and C.~Schmit,
\newblock Phys.Rev.Lett. {\bf 52}, 1 (1984);
T.~Seligman, J.~Verbaarschot, and M.~Zirnbauer,
Phys. Rev. Lett. {\bf 53}, 215 (1984);
T.~Seligman and J.~Verbaarschot, Phys. Lett. {\bf 108A} (1985) 183;
B.Simons, A. Szafer and B. Altschuler, {\it Universality in quantum chaotic
spectra}, MIT preprint (1993);  E. Br\'ezin and A. Zee,
Nucl. Phys. {\bf B402} (1993) 613; H. Weidenm\"uller,
\newblock in Proceedings of T. Ericson's 60th birthday.

\bibitem{MEHTA-1967}
M.~Mehta,
\newblock {\it Random Matrices}, Academic Press, New York, 1967.

\bibitem{MEHTA-1991}
M.~Mehta,
\newblock {\it Random Matrices}, Academic Press, San Diego, 1991.

\bibitem{DYSON-1962}
F.J. Dyson,
\newblock J. Math. Phys. {\bf 3} (1962) 140.

\bibitem{VERBAARSCHOT-ZAHED-1993}
J.~Verbaarschot and I.~Zahed, Phys. Rev. Lett. {\bf 70} (1993) 3852.

\bibitem{GRADSHTEYN-RHYZIK}
I. Gradshteyn and I. Ryzhik, {\it Tables of Series, Integrals and Products},
Corrected and enlarged edition (Academic, San Diego, 1980).

\bibitem{MEHTA-1962}
M. Mehta, Nucl. Phys. {\bf 18} (1960) 395.

\bibitem{GL} J.Gasser and H.Leutwyler, Phys. Lett. B {\bf 184} (1987) 83;
{\bf 188} (1987) 477.

\bibitem{MUIR-1933}
T. Muir, {\it A treatise on the theory of determinants}, 1933 (reprinted
by Dover, New York, 1960).

\end{thebibliography}

\vfill
\eject
\end{document}